\pdfoutput=1
\RequirePackage{ifpdf}

\documentclass[12pt,cits,hyper]{JINST}

\usepackage{booktabs}
\usepackage{setspace}

\newcommand{\degr}{\ensuremath{^{\circ}}}

\newcommand{\minitab}[2][l]{\begin{tabular}{#1}#2\end{tabular}}

\title{Optical Readout of a Two Phase Liquid Argon TPC using CCD Camera and THGEMs}
\author{K.~Mavrokoridis\thanks{Corresponding author}, F.~Ball, J.~Carroll, M.~Lazos, K.~J.~McCormick, N.~A.~Smith, C.~Touramanis, J.~Walker \\
University of Liverpool,
Department of Physics, Oliver Lodge Bld, Oxford Street, Liverpool, L69~7ZE, UK\\

E-mail: \email{k.mavrokoridis@liverpool.ac.uk}}

\abstract{

This paper presents a preliminary study into the use of CCDs to image secondary scintillation light generated by THick Gas Electron Multipliers~(THGEMs)
in a two phase LAr TPC. A Sony ICX285AL CCD chip was mounted above a double THGEM in the gas phase of a 40~litre two-phase LAr TPC with the majority
of the camera electronics positioned externally via a feedthrough. An Am-241 source was mounted on a rotatable motion feedthrough allowing the
positioning of the alpha source either inside or outside of the field cage.
Developed for and incorporated into the TPC design was a novel high voltage feedthrough
featuring LAr insulation. Furthermore, a range of webcams were tested for operation in cryogenics as an internal detector monitoring tool. Of the range
of webcams tested the Microsoft HD-3000~(model no:1456) webcam was found to be superior in terms of noise and lowest operating temperature.
\\
In ambient temperature and atmospheric pressure 1 ppm pure argon gas, the THGEM gain was $\approx$1000 and using a 1~msec 
exposure the CCD captured single alpha tracks.
Successful operation of the CCD camera in two-phase cryogenic mode was also achieved. Using a 10~sec exposure a 
photograph of secondary scintillation light induced by the Am-241 source in LAr has been captured for the first time.

}


\begin{document}

\section{Introduction}

\begin{figure}[t]
\begin{center}
\begin{tabular}{c}
	\includegraphics[width=.50\textwidth]{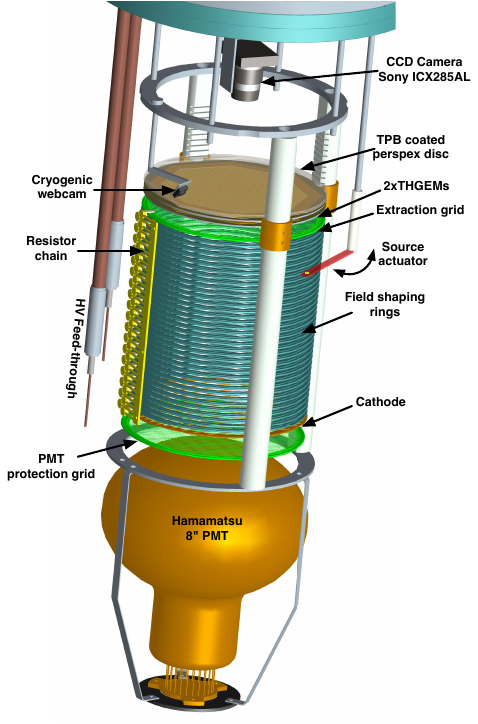}
\end{tabular}
\end{center}

\caption{A 3D CAD model of the internal detector assembly.}

\label{model}
\end{figure}

Two phase Liquid Argon~(LAr) Time Projection Chambers~(TPCs) that employ THick Gas Electron
Multipliers~(THGEMs,~\cite{Breskin:2009}) for charge readout and thereafter tracking 
reconstruction are being successfully
operated~\cite{Badertscher:3litre, Badertscher:2012} and are proposed to be installed in future giant two phase LAr detectors for 
Long Baseline Neutrino Physics~\cite{Rubbia:2004p3272, Rubbia:2009p3284}. The large number of charge readout channels required for THGEM operation 
and their unit price lead to a considerable cost if scaled up for multi-kilo-tonne LAr detectors.
Low energy experiments (dark matter, coherent  neutrino-nucleus scattering) and applications have different detector optimisation requirements. For
general reviews on noble liquid detectors see~\cite{Buzulutskov:2012, Chepel:2013}. 

Two-phase noble gas TPC technology was mainly developed for and utilised in dark matter detectors such as Zeplin-II~\cite{Zeplin2:2007}, XENON100~\cite{Xenon100:2012},
LUX~\cite{LUX:2013} and 
ArDM~\cite{Rubbia:2006, Laffranchi:2007p1321, Regenfus:2010p3319}.
The detection principle of such detectors relies upon the capture of 
prompt scintillation light~(S1) produced in the liquid phase and secondary delayed electroluminescence light~(S2) induced by the free-ionised electrons (generated during an interaction) as they are extracted into the gaseous phase.
When a THGEM is positioned in the gaseous phase, further secondary scintillation light production occurs in the THGEM holes as a result of electroluminescence.
When the electric field across the THGEM holes provides enough electron kinetic energy to produce an avalanche, 
the electroluminescence produced increases exponentially with THGEM electric field~\cite{Monteiro:2007, Monteiro:2008, Phil:2009, Monteiro:2012}.
Optical readout of THGEMs in two-phase argon detectors has been demonstrated using a Geiger-mode APD in~\cite{Phil:2009} for Vacuum Ultra Violet~(VUV) light, and
 in~\cite{Bondar:2010, Bondar:2012A, Bondar:2012B} for Near Infrared~(NIR).

Imaging of secondary scintillation light with a cryogenic CCD camera represents an 
exciting novel approach for track reconstruction since such an optical readout system is relatively simple and cost-effective.
Previous studies into the use of a CCD camera in combination with GEMs 
in room temperature gaseous argon mixture TPCs are reported in 
references~\cite{Fraga:2001, Fraga:2002, Fraga:2002b, Fraga:2003, Margato:2004}. Cryogenic studies in pressurised Ne gas using GEMs and an external CCD
camera have been presented in~\cite{Galea:2007}.


In this report we demonstrate successful detection of secondary scintillation light with a CCD camera mounted within the cryogenic environment of 
a two-phase argon detector for the first time.
This report addresses the challenges for the successful operation
of the CCD, setting the foundations for further research and development of this technology option.

\section{The Liverpool LAr Setup} \label{setup}

The internal detector assembly is housed within a 40~litre ultra high vacuum (UHV) chamber, suspended from the top 13.25-inch conflat (CF) flange. 
The top flange has a centrally positioned DN 100CF port surrounded symmetrically by four DN 40CF ports 
which allow the mounting of electrical and gas feedthroughs. A LAr recirculation/purification system is located on the side of this chamber.
To maintain cryogenic temperature the entire chamber sits within a 250~litre LAr open bath which itself is a vacuum jacketed vessel to minimise heat losses.
A detailed description of the cryogenic system and the novel recirculation system can be found in our previous publication~\cite{kostaspurity}.

An engineering CAD model of the internal detector assembly is shown in Figure~\ref{model}. 
An 8-inch Hamamatsu R5912-02MOD PMT is positioned at the bottom of the UHV chamber facing upwards towards the field cage.
The PMT is coated with tetraphenyl-butadiene (TPB) in order to shift the argon VUV scintillation light 
to 430~nm, well within the PMT's high quantum efficiency range~\cite{WLSkostas}. The PMT signal was recorded using an Agilent DP1400 digitiser.
Two THGEMs are positioned in the gaseous electron extraction region above the field cage and the CCD camera is placed 10~cm above
the THGEMs facing downwards. Also incorporated is a level monitor webcam positioned on the side of the extraction region 
and high voltage feedthroughs.

\subsection{The Field Cage}

The field cage creates a uniform electric field in order to drift ionisation electrons to the surface of the liquid volume. 
The field cage is composed of 38 178~mm diameter stainless steel field shaping rings (FSR).
Each FSR is 1.6~mm thick and positioned with an inter-ring spacing of 4~mm, thus a total drift length 
of 20~cm is created. 

The FSRs are supported by three macor rods\footnote{Macor is machinable glass ceramic type insulation material that has very low outgassing properties.}. 
Each FSR, the cathode grid and the extraction grid
are electrically connected with 100~M$\Omega$ resistors. These resistors are soldered to a PCB that was designed to have a structure resembling that 
of a flexible comb to compensate for materials with different Coefficients of Thermal Expansion~(CTE).
The connection to each FSRs is made via small clamps soldered to the tips of the PCB thus allowing for fast assembly/disassembly of the whole resistor chain.
At the top of the resistor chain is the extraction grid made out of a stainless steel mesh. The extraction grid and the bottom of the THGEM electrode define the 1 cm extraction 
region. In addition, the extraction grid and the THGEM serve as a parallel plate capacitor allowing monitoring of the liquid level while filling.
Below the cathode there is a grounded PMT protection grid.

The entire field cage is wrapped in 3M{\small\texttrademark}-foil coated with 1~mg/cm$^2$ TPB
via vacuum evaporation in order to increase the PMT light collection (not shown in Figure~\ref{model} for clarity).  Further information regarding
TPB coated reflector performance can be found in~\cite{WLSkostas}.

\subsection{LAr Insulated High Voltage Feedthrough}
\label{HV_feed}
A minimum absolute voltage of 20~kV needs to be supplied to the cathode in order to drift sufficient electrons 20~cm in LAr and to extract them to the gas phase.
LAr, as opposed to GAr, is an excellent insulator with an expected breakdown voltage value of 1.1~-~1.4~MV/cm~\cite{Swan:1961}. 
For this reason a novel feedthrough design was made that brings all the high voltage connections into the liquid phase.

The high voltage feedthrough consists of two commercial ceramic UHV feedthroughs each welded to the end of a  12~mm internal diameter stainless steel pipe.
Within each pipe a 30~kV coaxial cable is soldered to the feedthrough copper pin. 
The solder connection is covered with a 1~mm thick PTFE sleeve, which provides insulation for up to 10~kV in an air environment. 
In order to achieve the higher
insulation required for two-phase cryogenic operation, GAr is inserted into the pipes using Swagelok$^{\small\textregistered}$ fittings. 
Since the bottom of the pipes are immersed in LAr liquefaction takes place, thus creating the necessary insulation to the connections. 
This novel approach allows the system to recover if any discharge should occur as a result of, for example, gas bubbling.

\subsection {The THGEMs}

\begin{figure}[t]
\begin{center}
\begin{tabular}{c}
	\includegraphics[width=.80\textwidth]{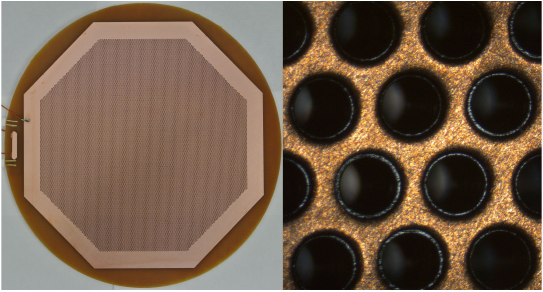}
\end{tabular}
\end{center}

\caption{A photograph of the THGEM on the left and a magnified region on the right, showing the 
50~$\mu$m dielectric rim.}

\label{tgemphotol}
\end{figure}

For all measurements reported in this paper, two THGEMs manufactured by the CERN TS/DEM workshop were used~\cite{DEMworkshop}.
The amplification region of the THGEMs has an octagonal shape with a 150 cm$^2$ surface area.
Within this region there are approximately 23000 holes that have been mechanically drilled with standard PCB techniques in a copper cladded glass epoxy plate.
The copper extends 1~cm from the perimeter of the amplification region for optimal shaping of the electric field at the edges of the active volume.
Each THGEM is 1~mm thick and and each hole has a diameter and pitch of 500~$\mu$m and 800~$\mu$m respectively. 
A 50~$\mu$m dielectric rim is also etched around each hole to extend the breakdown voltage of the THGEM.
A photograph of the THGEM is shown in Figure~\ref{tgemphotol}. The optical transparency of the THGEM is 35\%.
When mounted in the detector the space between the two THGEMs is 4~mm and care was taken to align the holes of the top and bottom THGEM.

\subsection{Charge Readout and Source Selection}
\label{source}
A 30~kBq Am-241 alpha source was used to calibrate the detector in gas and liquid operation. 
Alpha tracks deposit all their electrons within 4~cm in pure argon gas and therefore Am-241 is an 
ideal source for gas measurements and optical imaging with the CCD camera.  Additionally, 
the energy deposition of alphas in gas compares to a 
muon deposition energy in LAr which is $\approx$2.3~MeV/cm~\cite{ArgoNeuT:2012}.
The source was mounted facing upwards  3~cm below the extraction grid using a rotation motion feedthrough that allows the source to enter and exit the field cage through the gap between two FSRs. 
As alphas do not penetrate the field cage from outside, we can effectively switch on and off the source ionisation signal.
For the two phase run an external Cs-137 high rate gamma source was also used.

The gain of the THGEMs is defined as the ratio of the charge produced after amplification over the initial charge produced before amplification.
No corrections are incorporated for electron losses due to grid transparency and electron recombination. 
In LAr the 5.4 MeV alphas are expected to produce on average 228800 primary electron-ion pairs whereas in GAr 204500 electron-ion pairs will be produced
assuming a W-value of 23.6~eV and 26.4~eV respectively~\cite{Miyajima_ionpair}.
The Am-241 source also produces gammas which are predominantly at 59.5~keV and these are expected to give rise to approximately 2500 electron-ion pairs in LAr.
The gain of the THGEMs was measured using an ORTEC 142IH preamplifier connected to the top THGEM electrode. This preamplifier is calibrated by the manufacturer 
to produce 0.05~$\mu$V per electron-ion pair. 
For each preamplifier amplitude measurement the average of 1000 pulses was calculated using the mathematical function on a Lecroy 9374TM oscilloscope.

\subsection{Characterisation of the CCD Camera at Low Temperature}

The CCD camera used for all measurements was an Artemis FS14 containing a Sony ICX285AL chip. The camera characteristics
are summarised in Table~\ref{ccd_manufacture}. This chip is popular in the astronomy community because of its good low light detection performance and relatively low cost. 
In our case stable operation down to $-185$~$\degr$C is also critical.

Although it is common practise to cool down camera chips to between $-20$~$\degr$C and $-80$~$\degr$C for noise minimisation, the manufacturer of the Sony ICX285AL chip will only 
guarantee operation down to $-10$~$\degr$C~\footnote{The manufacturers True Sense Imaging and e2v produce CCD chips guaranteed to operate down to  $-60$~$\degr$C and $-200$~$\degr$C respectively.}. 
Additionally, the camera as a whole system (including the digitiser card) will not operate at such low temperatures without the development of cold electronics.
We circumvent this issue by mounting the majority of the electronics (i.e. the digitiser) externally, connected to the CCD sensor via a custom made Kapton cable
and feedthrough.

The chip's response to low temperature was tested before assembly in the detector using a separate apparatus. 
A thermocouple was attached to the back of the chip which was then was cooled down at a rate of about 3~$\degr$C/min.
The chip thermal noise reduced significantly during the cool-down, however, the chip stopped functioning below $-120$~$\degr$C.
The CCD gain and the Read Out Noise~(RON) for 28~$\degr$C and $-100$~$\degr$C were measured and are shown in Table~\ref{ccd_measurements}. 
To allow chip operation in
a colder environment a resistor type heater was made and placed on the back of the chip. 
This allowed very nice heat exchange control of the chip which, as a result, operated down to $-190$~$\degr$C.
In two-phase operation the chip was typically kept at $-60$~$\degr$C.

As the secondary scintillation light produced in the THGEM holes is mainly in the VUV region, special care has to be taken with the optics used.
VUV grade lenses are very expensive and additionally VUV is out of the spectrum range of the CCD. An obvious way around this issue is to
coat a normal camera lens with wavelength shifter (WLS), however, this was inadequate for this setup.
As an alternative we coated a transparent disk with WLS and placed this directly above the THGEM, thus VUV light is converted to visible before it reaches the camera lens.
Specifically, a 178~mm diameter perspex disk coated with 0.05~mg/cm$^2$ TPB was positioned 4~mm above the top THGEM. As VUV light doesn't penetrate perspex, the coated side 
of the disk was placed facing downwards.
The lens used in the setup is a Fujinon DF6HA-1B which has a small focal ratio of f/1.2 allowing more light to reach the chip. In order to mount this lens onto the chip,
a CCD chip holder with a C-mount thread was manufactured.

It is also worth mentioning that it was found that small impurities
in GAr, even on the 40~ppm level (based on argon scintillation slow component decay time measurements~\cite{kostaspurity}), coming from detector component outgassing 
is enough to produce visible light within the spectrum range of the CCD thus wavelength shifter is redundant in this case.
However, for higher purity levels the signal is visible only with the use of WLS.

\begin{table}
\caption[Sony CCD Characteristics.] {Artemis FS14 CCD  PCB assembly characteristics.}

\begin{center}
\vspace{4mm}
{\footnotesize \begin{tabular}{lc}
\toprule
\multicolumn{2}{c}{CCD camera Artemis FS14} \\
\midrule
CCD sensor type: & Sony ICX285AL\\ 
CCD sensor design: & Monochrom, Progressive scan, Interline transfer\\
Sensor dimensions: & 8.98~mm$\times$ 6.71~mm, Diagonal 11.21~mm, 4:3, Type 2/3" \\
Pixel resolution (H$\times$V): & 1391$\times$1039, 1.45 Megapixels \\
Pixel size: &6.45~$\mu$m~$\times~$6.45~$\mu$m \\
Full well capacity: & 17.500 e$^-$\\
Typical Gain (temp dependant): & 0.267 e$^-$/ADU \\
Read out noise (temp dependant): & 3.7~e$^-$ \\
Spectrum range: & 300~nm - 1050~nm \\
Quantum efficiency at 430~nm: & 50 \% \\
Min exposure time: & 1~msec \\
Dynamic range: & 1:4730 \\
ADC and data format: & 16 bit, RAW Fits \\
Binning: & 1$\times$1, 2$\times$2, 3$\times$3, 4$\times$4, 5$\times$5, 6$\times$6, 7$\times$7, 8$\times$8 via software \\
Data interface: & USB 2.0 \\
\bottomrule

\end{tabular}}
\end{center}
\label{ccd_manufacture}
\end{table}

\begin{table}
\caption[Sony CCD Characteristics.] {Sony ICX285AL CCD chip gain and Read Out Noise (RON) measurements.}

\begin{center}
\vspace{4mm}
{\footnotesize \begin{tabular}{cccc}
\toprule
 \minitab[c]{CCD Gain\\ at 28~$\degr$C~(e$^-$/ADU)}
&\minitab[c]{CCD Gain \\ at $-100$~$\degr$C~(e$^-$/ADU)} & \minitab[c]{CCD RON \\ at 28~$\degr$C~(ADU)}
&\minitab[c]{CCD RON \\ at $-100$~$\degr$C~(ADU)}\\
\midrule
 0.32  & 0.27 & 28.30 & 16.50 \\

\bottomrule
\end{tabular}}
\end{center}
\label{ccd_measurements}
\end{table}

\subsection{Cryogenic Webcams for Monitoring of LAr Level and Discharges}

The use of monitoring webcams in LAr TPCs is an original concept that has not yet been explored. 
Such cameras provide an invaluable insight into the internal workings of the detector, allowing monitoring of the LAr level (which in a two-phase system must be precisely maintained within the extraction
region) as well as identification of discharge sources in the volume and heat-load inputs that cause bubbling and therefore may induce discharges.
Before these benefits can be realised, cameras that can operate within LAr must be identified. 

A series of tests were performed on very low budget commercial webcams in order to identify those that
can operate down to LN$_{2}$ temperature ($-197$~$\degr$C). 
The operation of 15 webcams was tested during 30 minutes of immersion in LN$_{2}$, and the Microsoft VX-1000, Microsoft
VX-3000 and Microsoft HD-3000 models were found to work fairly well under these conditions with the HD-3000 having the least noise.
It should be noted that all webcams tested functioned 
down to $-100$~$\degr$C. All webcams identified to function in LN$_{2}$ have a CMOS chip. The dimensions of these cameras after their outer casing is removed are
advantageously small allowing for positioning within the confined spaces in the detector.
Unfortunately, the manufacturer provides minimal technical information for these cameras, for example the CMOS chip model is not specified.

An extra feature of the HD-3000 model in comparison to the Microsoft VX-1000 and VX-3000 is that it has an auto focus which impressively allows one to focus sufficiently outside or inside the liquid. 
In order to assure result reproducibility three webcams of each model were purchased and all identical models were shown to have similar performances. 
However, it was found that the HD-3000 models produced since the year 2011 (model no:1492) do not work in cryogenics and 
therefore care should be taken to purchase the earlier model (model no:1456).
A summary of the identified cryogenic webcam specifications is shown in Table~\ref{table:webcams}.
As a result of this study the Microsoft HD-3000 webcam (2010 batch) was positioned in the detector along with four LEDs that light up the volume.


\begin{table}
\caption[webcams.] {Characteristics of the successful cryogenic web cameras. }

\begin{center}
\vspace{4mm}
{\footnotesize \begin{tabular}{lcccc}
\toprule
 \minitab[c]{Webcam\\ Name}
&  \minitab[c]{Model \\ Number} & Sensor Type  & Focus
&Comments \\
\midrule
 Microsoft VX-1000  & 1080 & CMOS & Manual& -\\
Microsoft VX-3000  & 1076 & CMOS & Manual& -\\
Microsoft HD-3000  & 1456 & CMOS & Auto&  \minitab[c]{Model no:1492 \\doesn't work in cryogenics} \\

\bottomrule
\end{tabular}}
\end{center}
\label{table:webcams}
\end{table}

\section{Gaseous Ambient Temperature Operation}
\label{gastest}
Prior to operation in LAr, various tests in gaseous pure argon were made in order to understand
the THGEM gain performance and the camera sensitivity. 
As such, for various THGEM bias voltages the THGEM gain, PMT light collection, and CCD intensity were monitored.
All the CCD images were recorded in a FIT format and they were analysed using IDL~\cite{idl}. 

Before filling, the chamber was
evacuated down to 6$\times$10$^{-7}$~mb in order to achieve minimum outgassing from the detector components.  Furthermore,
all sets of measurements were taken within a time span of 3 hours in order to ensure a stable argon purity.

\begin{figure}[t!]
\begin{center}
\begin{tabular}{c}
	\includegraphics[width=.7\textwidth]{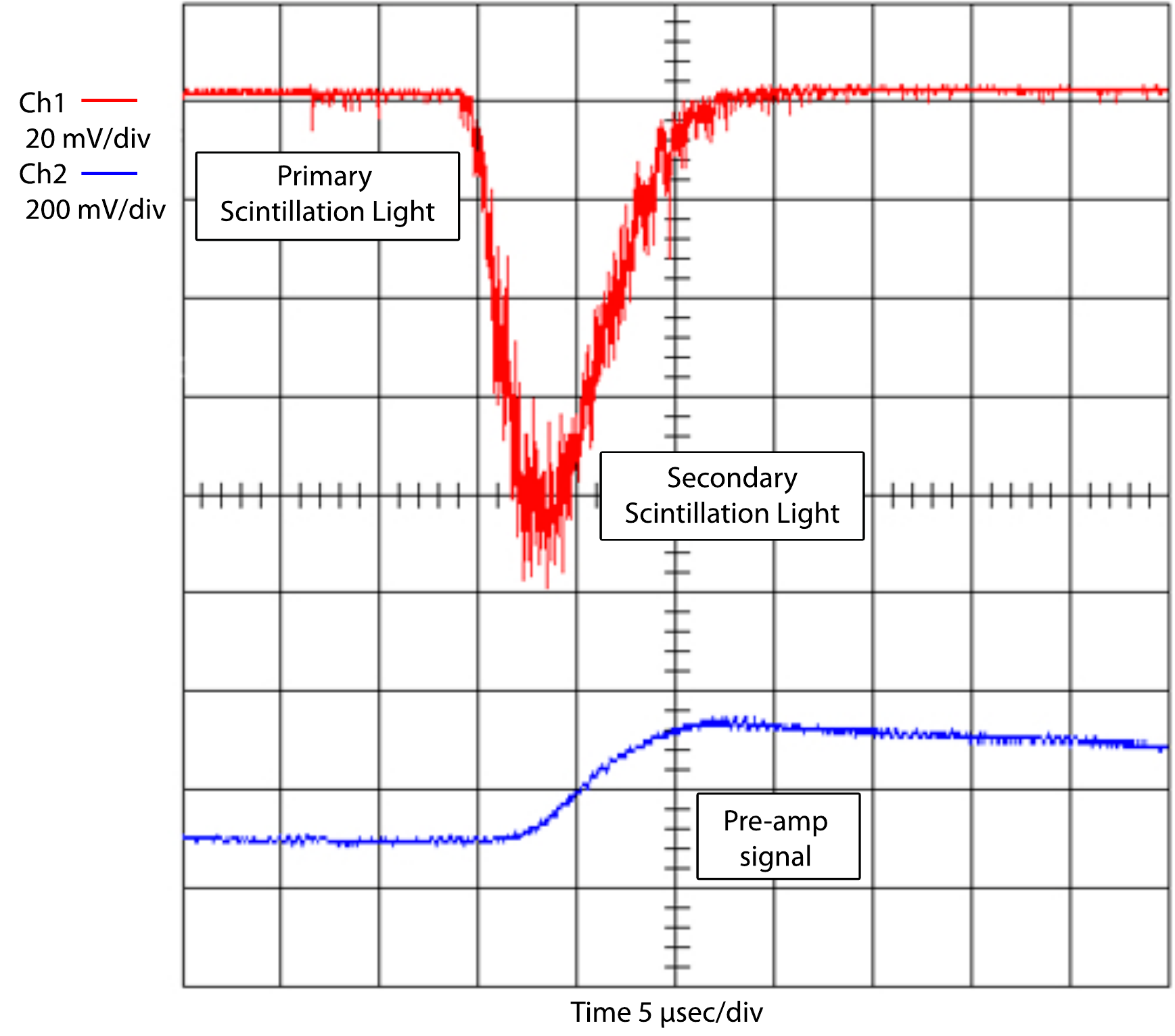}
\end{tabular}
\end{center}

\caption[eventsample]
{ A sample event of primary and secondary scintillation light with the corresponding charge signal from the preamplifier in pure 1~ppm argon gas. The THGEM gain was $\sim$20.}

\label{eventsample}
\end{figure}

\begin{figure}[b]
\begin{center}
\begin{tabular}{c}
	\includegraphics[width=.62\textwidth]{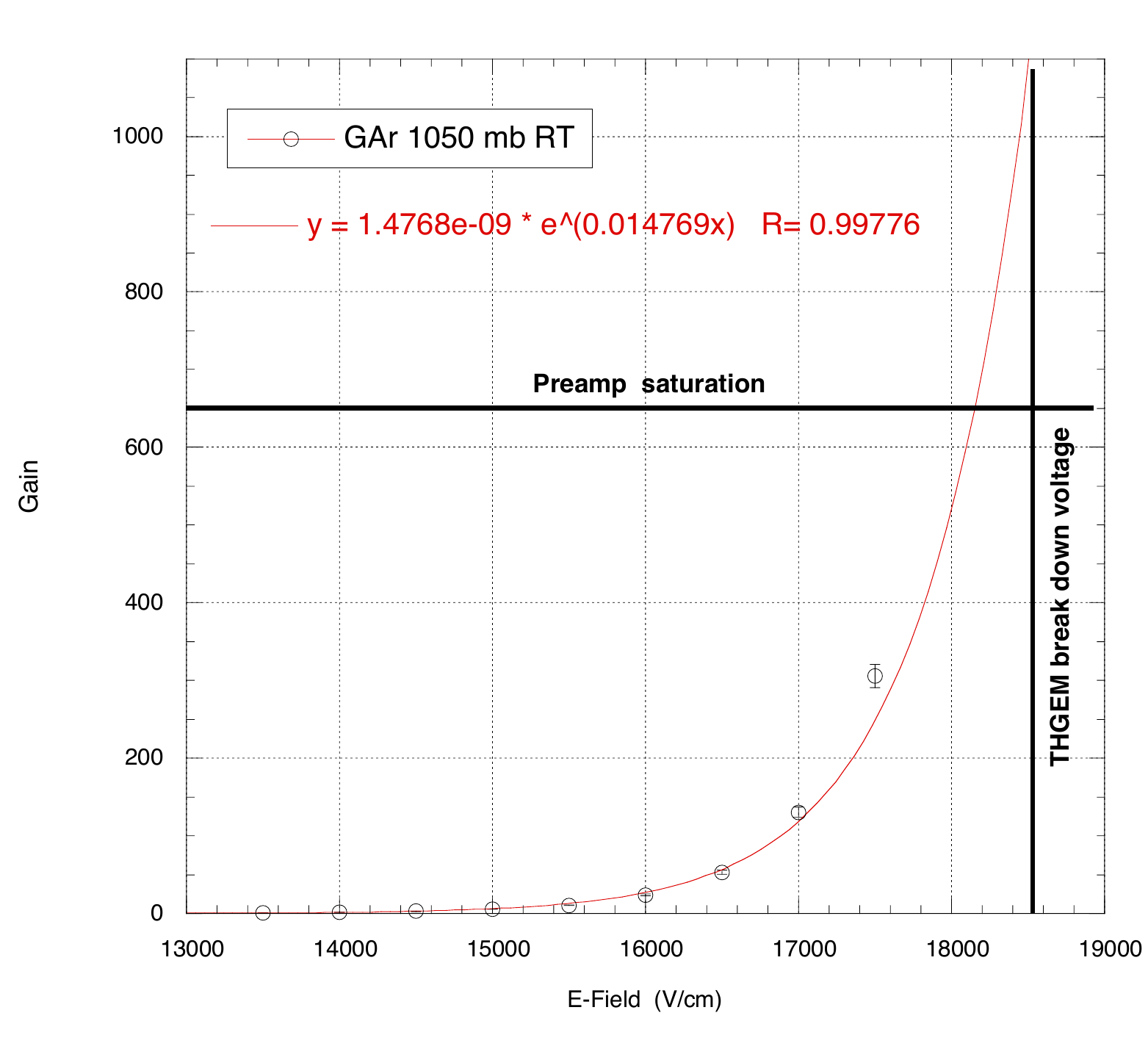}
\end{tabular}
\end{center}

\caption{Gain variation with THGEMs field. Gain measurements above 18~kV/cm were not possible as the pre-amp signal was saturated. The break down voltage of the THGEMs was approximately 1850~V.}

\label{gainvsERT}
\end{figure}

The electric field configurations for the gas tests are presented in Table~\ref{GAr_Efields}.
A sample of pure argon gas primary and secondary scintillation light and the charge signal from the preamplifier due to an alpha event are 
shown in Figure~\ref{eventsample}.
The overall gain from the top THGEM as a function of bias voltage applied is shown in Figure~\ref{gainvsERT}.
The preamplifier saturated at
THGEM electric fields higher than 18~kV/cm, therefore the maximum gain based on extrapolation of the exponential fit is approximately 1000.
The minimum amount of charge that the preamplifier can distinguish from the noise was determined to be 175000 electrons, and this corresponds 
to a gain of 0.85~(due to electron losses).

\begin{figure}[b]
\begin{center}
\begin{tabular}{c}
	\includegraphics[width=.62\textwidth]{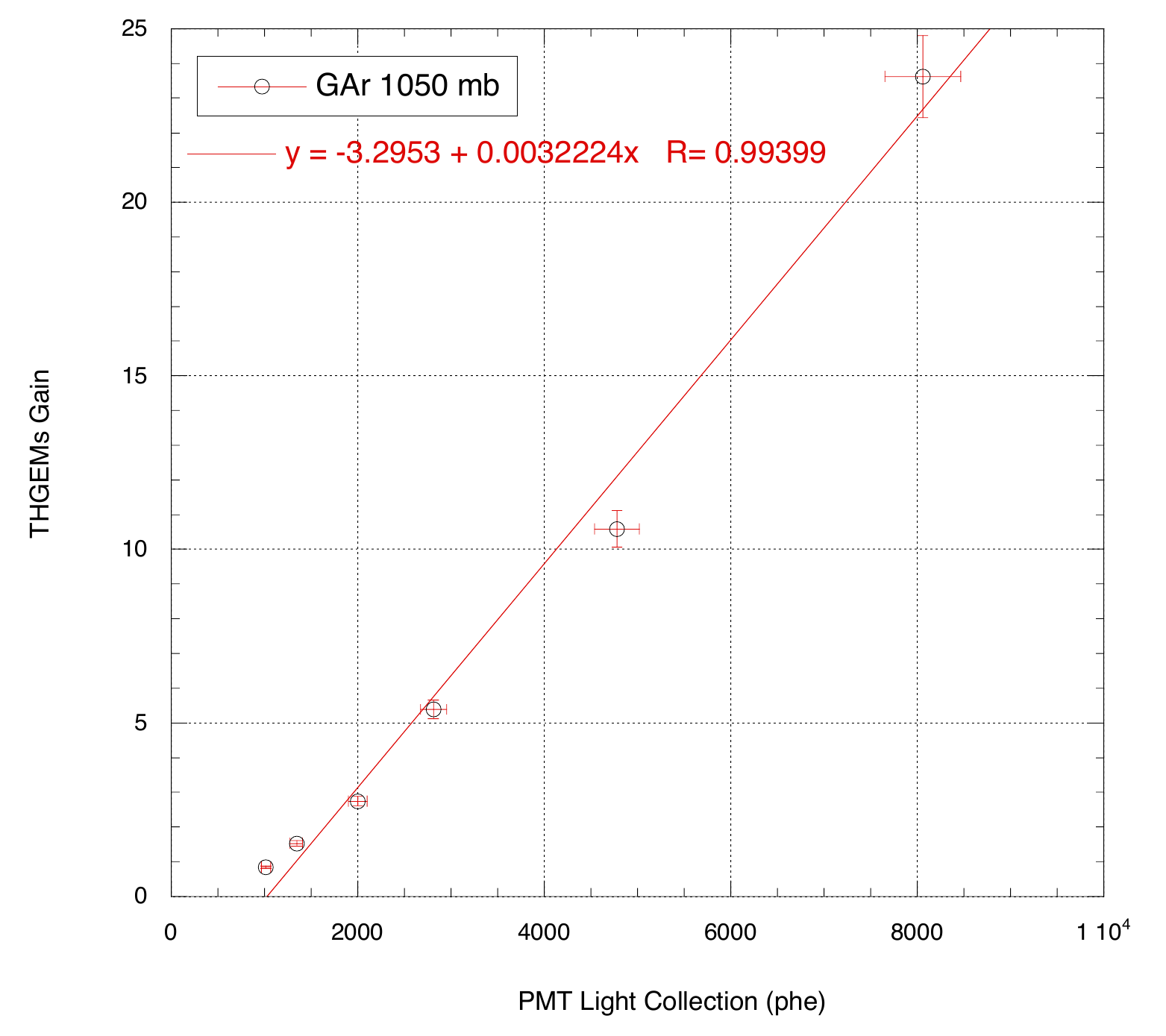}
\end{tabular}
\end{center}

\caption{Correlation between THGEMs gain and PMT light collection.  For THGEMs gain values higher than 25 the PMT was saturated and therefore no data are shown, although the highest gain in gaseous argon at ambient temperature was approximately 1000.}

\label{gainvsPMTlight}
\end{figure}

\begin{figure}[t]
\begin{center}
\begin{tabular}{c}
	\includegraphics[width=.65\textwidth]{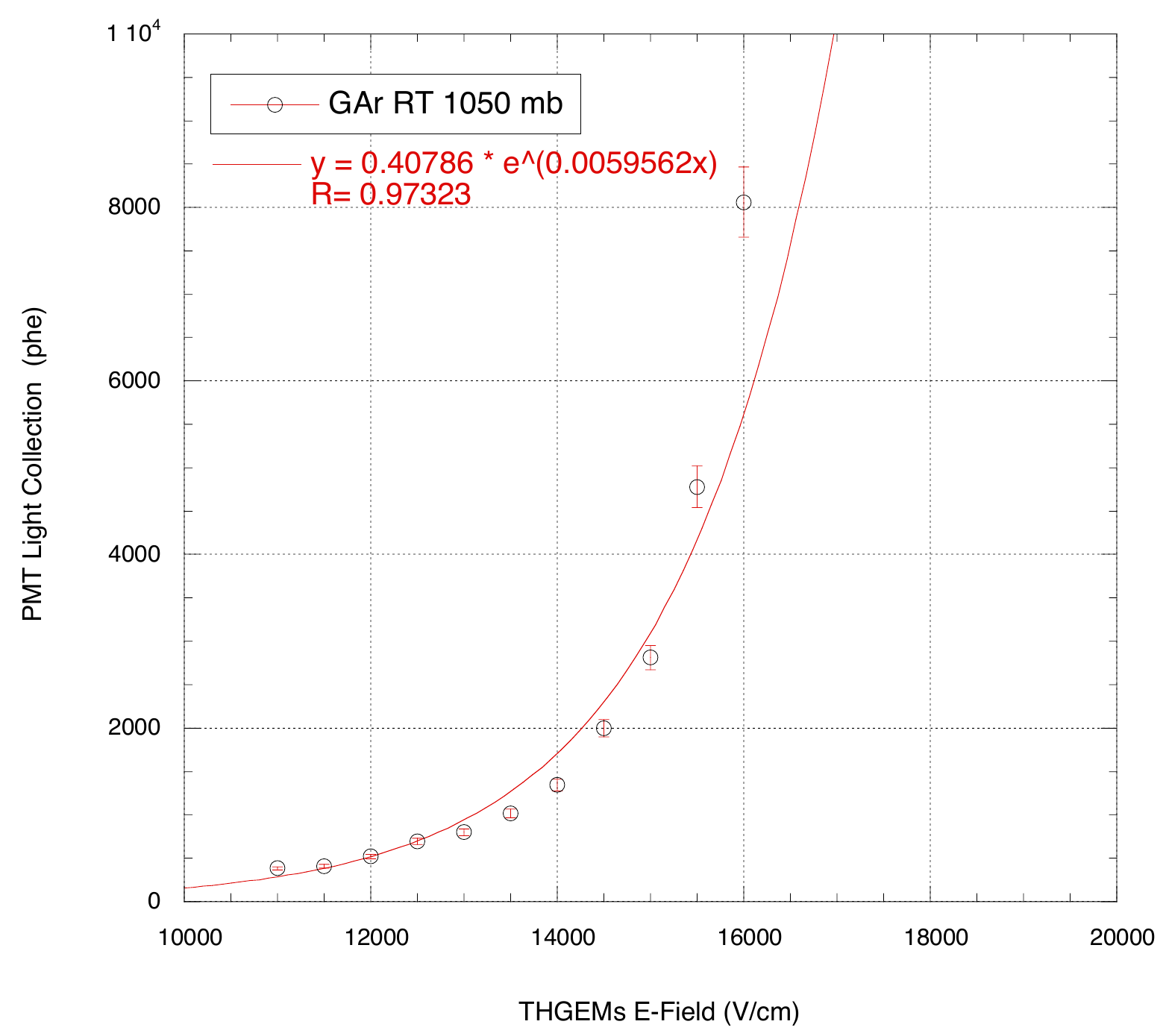}
\end{tabular}
\end{center}

\caption{ Variation of PMT light collection with THGEMs field. The PMT was saturated for fields higher than 16~kV/cm.}

\label{PMTlightTGEMfield}
\end{figure}

The PMT light collection increase with THGEM gain is shown in Figure~\ref{gainvsPMTlight} and approximately follows 
a linear relationship. Data for THGEM gains higher than 25 are not shown as the light production was very high and the PMT saturated.  
Figure~\ref{PMTlightTGEMfield} shows an exponential relationship between PMT light collection and THGEM electric field.

\begin{figure}[t]
\begin{center}
\begin{tabular}{c}
	\includegraphics[width=.65\textwidth]{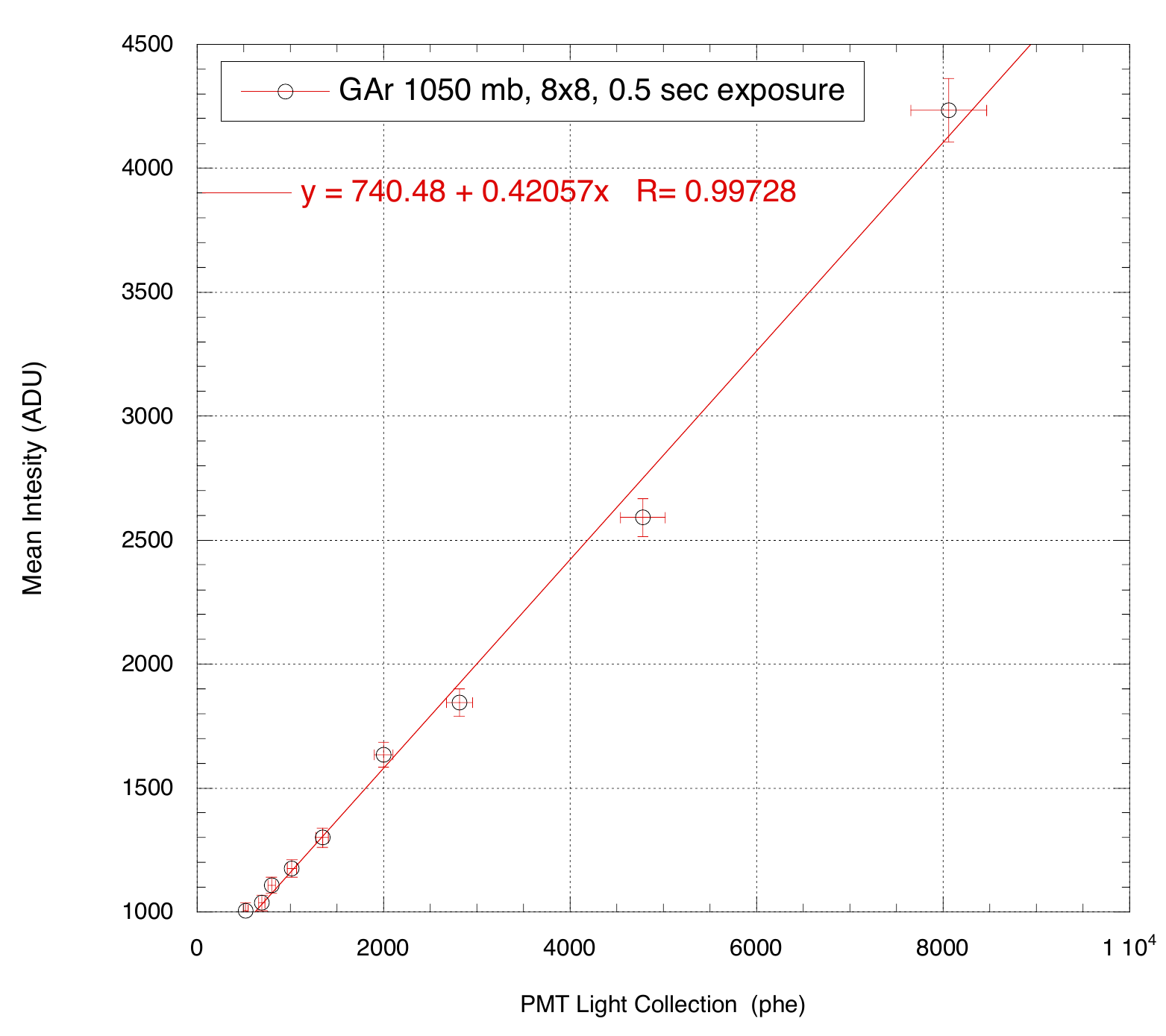}
\end{tabular}
\end{center}

\caption{ Correlation between PMT and CCD light collection. The mean intensity of the CCD refers to the Gaussian mean value from the
image region that contains the alpha source.}

\label{PMTlightvsCCD}
\end{figure}

A comparison between the light collected from the PMT and the CCD camera is shown in Figure~\ref{PMTlightvsCCD}.
The mean intensity of the CCD camera was calculated by finding the Gaussian mean of the
intensity histogram for the pixels containing the alpha source region. 

\begin{figure}[t]
\begin{center}
\begin{tabular}{c}
	\includegraphics[width=.65\textwidth]{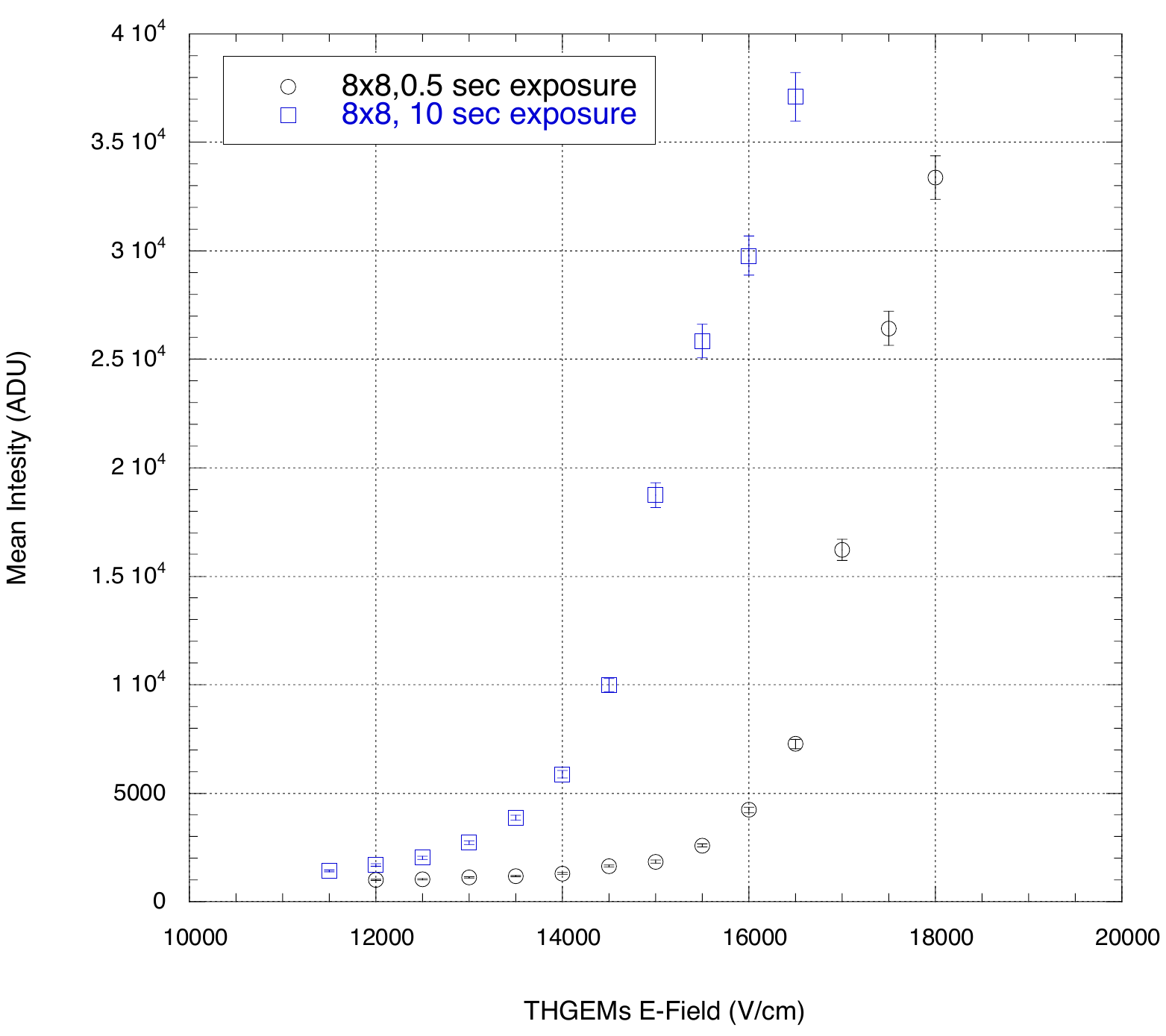}
\end{tabular}
\end{center}

\caption{ CCD mean intensity variation with THGEMs field for 0.5 and 10~sec exposure.}

\label{meanInt_vs_E}
\end{figure}

\begin{figure}[t]
\begin{center}
\begin{tabular}{c}
	\includegraphics[width=.65\textwidth]{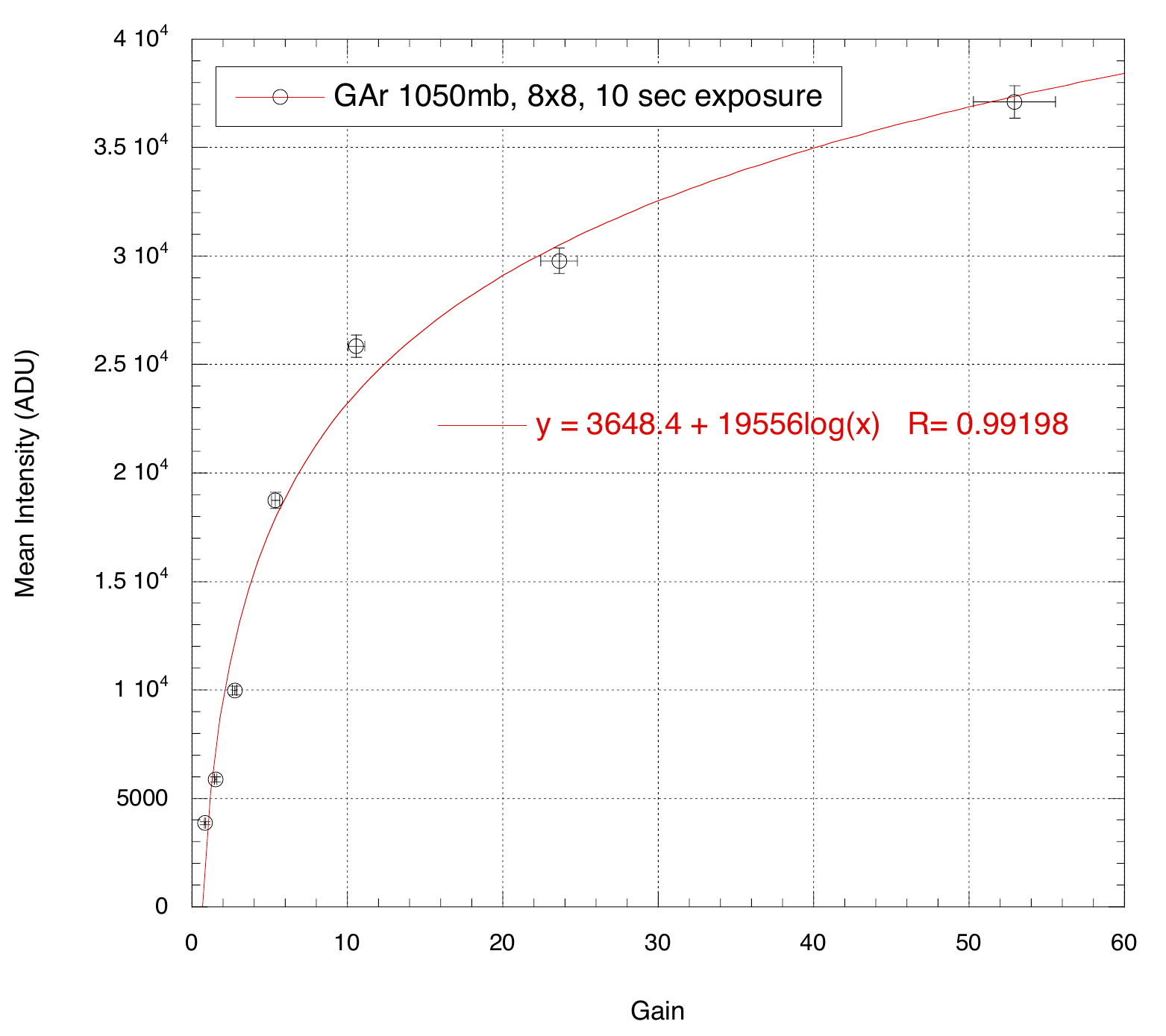}
\end{tabular}
\end{center}

\caption{ Correlation between CCD intensity and THGEMs gain. A gain of 1 corresponds to approximately 4000~ADU for a 10~sec exposure.}

\label{ccdinte_vs_gain}
\end{figure}

The effects of the electric field and the gain on CCD intensity were also measured and are shown in Figure~\ref{meanInt_vs_E} and Figure~\ref{ccdinte_vs_gain} respectively.

For 10 sec exposure using 8$\times$8 binning the minimum THGEM gain at which the CCD camera can detect light is
 approximately 1 and the maximum THGEM gain before CCD saturation is  approximately 53.

\begin{figure}[t]
\begin{center}
\begin{tabular}{c}
	\includegraphics[width=.95\textwidth]{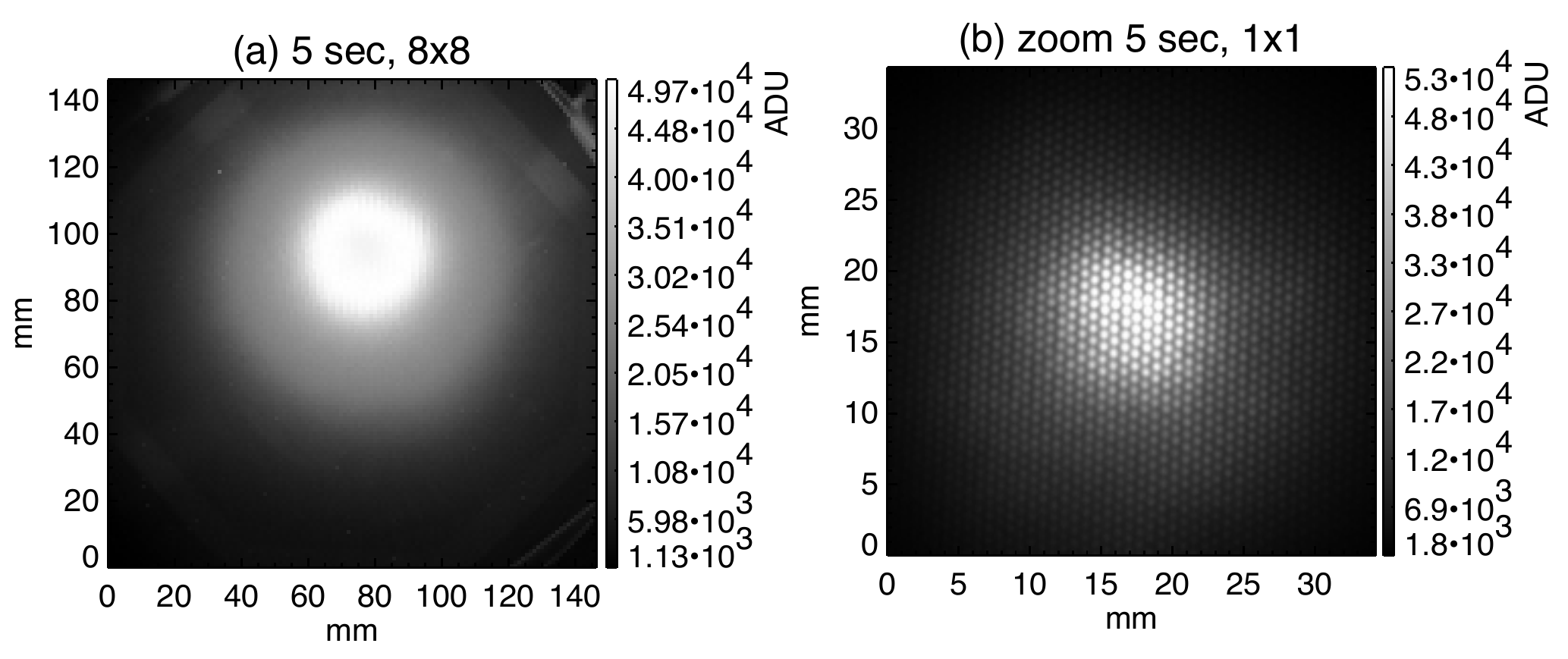}
\end{tabular}
\end{center}

\caption{Images of the secondary scintillation light in ambient temperature and pure argon gas induced by Am-241 for a THGEM gain of 600. 
a) 8$\times$8 binning and 5~sec exposure, illumination of the whole THGEM plane. 
b) A zoom of the alpha source region at a high 1$\times$1 binning resolution and 5~sec exposure, the individual THGEM holes are clearly visible.}

\label{zoom8bin1binfinal}
\end{figure}

\begin{figure}[t]
\begin{center}
\begin{tabular}{c}
	\includegraphics[width=.85\textwidth]{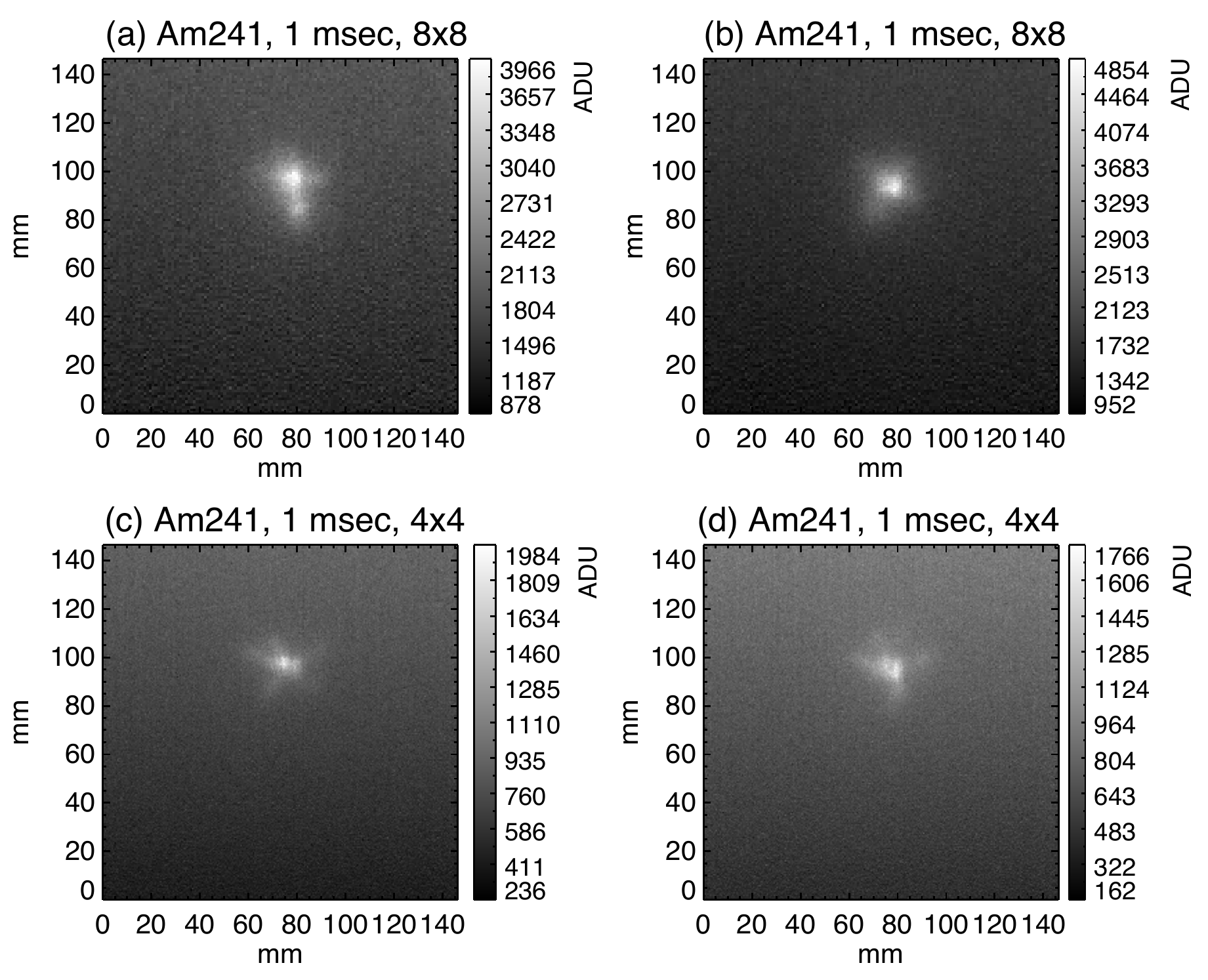}
\end{tabular}
\end{center}

\caption[1msec alpha tracks gallery.]
{ A gallery of alpha tracks in pure argon gas and ambient temperature. The electric field for both THGEMs was set to 18.5~kV/cm  and the gain was approximately 
1000. The top images were captured with 8$\times$8 binning whereas the bottom ones with 4$\times$4 binning.}

\label{1msec4bin8bin}
\end{figure}

CCD photographs taken with a total THGEM gain of 600 using a 5~sec exposure are shown in Figure~\ref{zoom8bin1binfinal}. The light collected using a 5 sec exposure and 8$\times$8 binning 
 is enough to light up the entire THGEM plane. For the same exposure but at a high 1$\times$1 binning resolution the individual 500~$\mu$m THGEM holes are clearly visible.
 For the minimum CCD exposure of 1~msec and for an approximate total THGEM gain of 1000 individual alpha tracks in pure argon can be identified.
 A gallery of these alpha tracks for 4$\times$4 and 8$\times$8 binning is shown in Figure~\ref{1msec4bin8bin}. The minimum THGEM gain required to separate
 individual alpha tracks at 1~msec exposure and 8$\times$8 binning is approximately 600.


\begin{table}[t]
\caption[Electric Field Configuration] {Configuration of the electric fields applied in room temperature
 gas operation.}

\begin{center}
\vspace{4mm}
{\footnotesize \begin{tabular}{lcccc}
\toprule
 & \minitab[c]{Distance to the \\ stage above (cm)} 
& Potential (kV) & \minitab[c]{Field to the stage \\ above (kV/cm)}\\
\midrule
THGEM$_{2}$ (top electrode) & - & +1.50 to +1.85  & -  \\
THGEM$_{2}$ (bottom electrode) & 0 &  0 & 15.0 to 18.5 \\
THGEM$_{1}$ (top electrode) & 0.4 & 0 & 0 \\
THGEM$_{1}$ (bottom electrode) & 0.1 & -1.50 to -1.85 & 15.0 to 18.5  \\
Extraction grid & 1.0 & -2.0  &  0.15 to 0.5  \\
Cathode & 20 & -4.0  & 0.1  \\

\bottomrule
\end{tabular}}
\end{center}
\label{GAr_Efields}
\end{table}


\clearpage

\clearpage

\section{Two Phase Operation}

After successful operation and characterisation of the THGEMs
 and CCD camera in room temperature, tests in cryogenic two phase conditions were performed.

The electron drift velocity in LAr at 0.5~kV/cm electric field is 1.6 mm/$\mu$s~\cite{Walkowiak:2000, Icarus:2004} thus for the entire drift volume electrons will take 125~$\mu$s to reach the surface, whereas  the electrons generated from the alpha source positioned 3~cm below the extraction grid will take 18.75~$\mu$s.
 The lifetime~($\tau$) of the drifting electrons is highly dependant on LAr purity and 
 can be approximated as a function of  O$_{2}$ equivalent impurity concentration ($\rho$)~\cite{Icarus:2004, Buckley:1989, Aprile:1985} as
$\tau$ [$\mu$s]$\approx$300/$\rho$ [ppb]. 
Therefore, a LAr purity level better than 2~ppb and 15~ppb is required for the drifting electrons to transit 20~cm and 3~cm respectively.

High purity is mainly achieved with 
 the recirculation/purification system which was detailed in~\cite{kostaspurity}. Nevertheless, great care was taken to start with a good initial LAr purity by
 minimising the outgassing of the internal detector components before highly purified argon 
 gas\footnote{An SAES getter model MicroTorr MC400 in conjunction with a cartridge containing a mixture of 3A, 4A and 13X molecular sieves were used.}
  was inserted into the detector. This was achieved by pumping the detector down 
 for a week to 6$\times$10$^{-7}$~mb vacuum and then pumping was continued into the cool-down stage until $-5$~$\degr$C was reached.
 In this way an even higher vacuum (1$\times$10$^{-7}$~mb) was achieved and any water impurities were frozen in the vessel.

 Filling of the detector by argon gas liquefaction lasted approximately 12~hours and during this period the performance of the
 THGEMs in cold gas was evaluated. The Am-241 source remained in the gas phase until the last stage of filling as it
 was positioned at a high level within the field cage. The thermocouple behind the CCD chip gave an estimate of the gas temperature
 in the vicinity of the THGEMs.

The LAr level was adjusted to half way between the extraction grid and the bottom THGEM by measuring the capacitance and by looking
within the detector volume with the cryogenic webcam. The webcam also revealed that the LAr level
 remained constant and steady during recirculation which allowed for data collection whilst the purification pump was on.

Prior to setting the cathode to a high voltage, gas argon was condensed in the pipes of the HV feedthrough to provide extra insulation as was described in Section~\ref{HV_feed}. The electric field configuration for the two phase operation is shown in Table~\ref{LAr_Efields}. 
It is well known that a 3~kV/cm electric field within the liquid phase is sufficient to extract the charge across the liquid gas
 boundary~\cite{Bolozdynya:1999, Dolgoshein:1970, Borghesani:1990}.

 It should also be noted that in LAr for 0.5~kV/cm drift field approximately 30\%, 40\% and 99\% of the charge induced by MIPs, gammas and alphas respectively will
 recombine before reaching the surface~\cite{Amoruso:2004recomb}. 
 Therefore, the total expected amount of electrons from the Am-241 source (alphas plus gammas) that will reach the surface of the
 liquid before they recombine will be approximately 4000 considering the W-value of LAr discussed in Section~\ref{source}.
 
\begin{figure}[t]
\begin{center}
\begin{tabular}{c}
	\includegraphics[width=.95\textwidth]{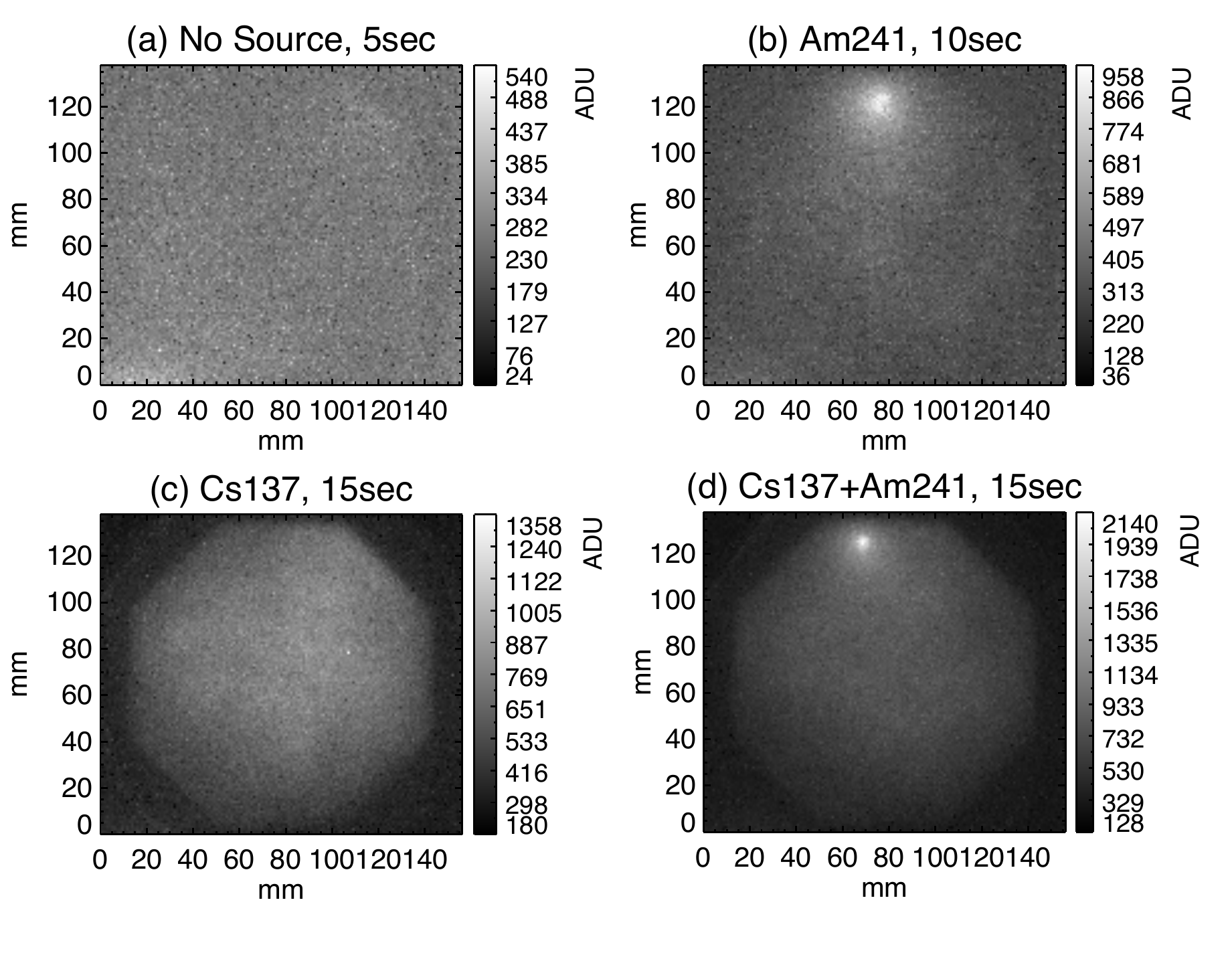}
\end{tabular}
\end{center}

\caption{Images of the top THGEM in cryogenic two phase operation a) with no source, b) with the Am-241 source placed within the active region, 
c) with only the external Cs-137 source, d) with both the Am-241 source within the active region and the external Cs-137 source. For all four images the bottom THGEM field was set to 40~kV/cm while the top was set to 20~kV/cm. The gain of the THGEMs was estimated to be $\lesssim$~45 and the binning for all images was 8$\times$8.}

\label{cryophotos}
\end{figure}

\begin{figure}[t]
\begin{center}
\begin{tabular}{c}
	\includegraphics[width=.74\textwidth]{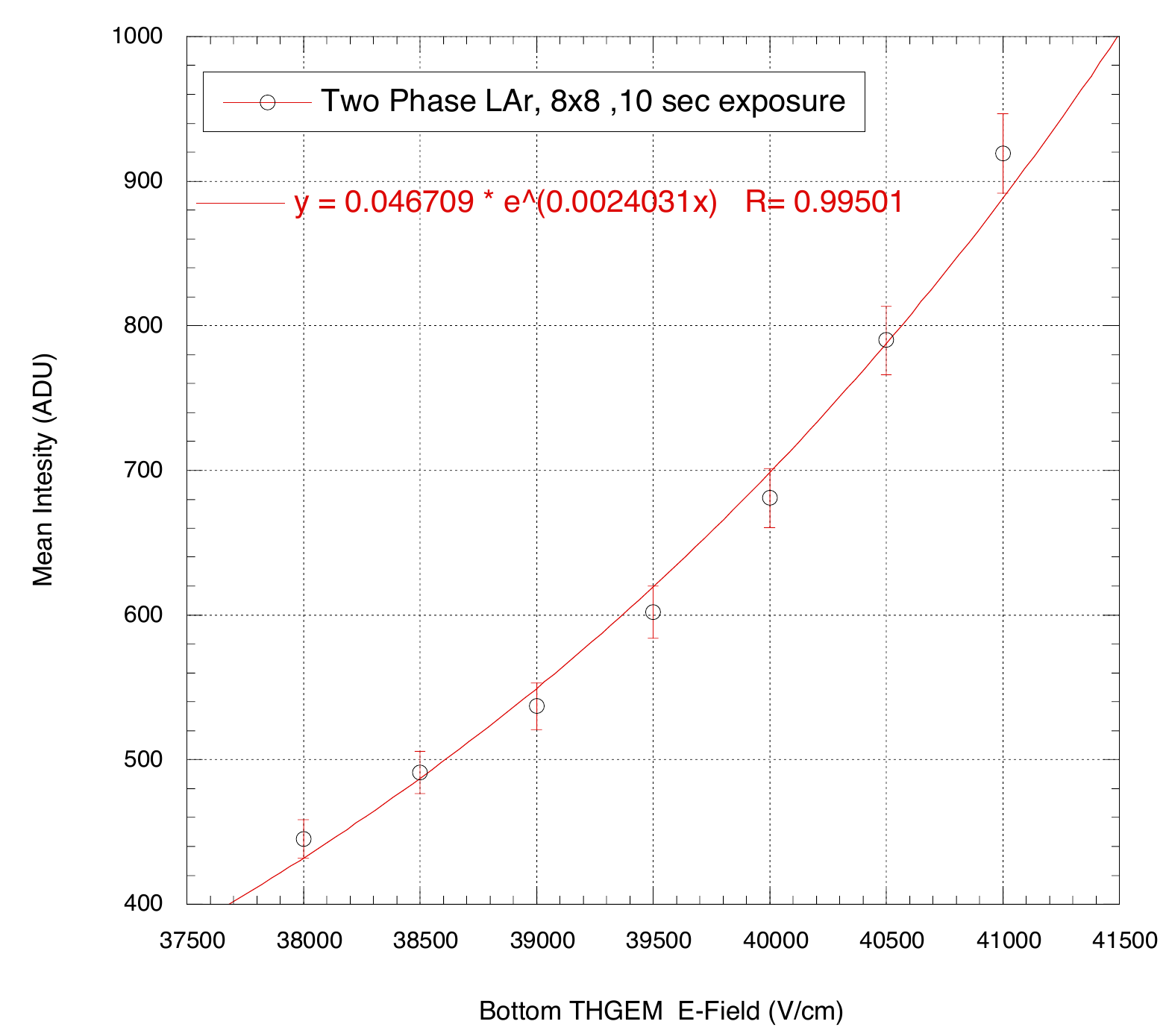}
\end{tabular}
\end{center}

\caption[Two phase CCD intensity data]
{ Variation of CCD intensity with bottom THGEM field in a two phase operation. The exposure was set to 10~sec and the binning was 8$\times$8.}
\label{LArCCDIntevsEfield}
\end{figure}

Extraction of electrons produced by the Am-241 source was verified by observing
the secondary electroluminescence light produced in the extraction region with the PMT.  
Further light production in the THGEM holes was achieved by increasing the electric field in the THGEMs. The maximum electric fields in the bottom and top THGEMs before discharges occurred were
 41500~volts/cm and 22000~volts/cm respectively. With such electric fields we have captured the first images of an Am-241 source submerged in LAr using a CCD camera
 and 10 to 15~sec exposures as shown in Figure~\ref{cryophotos}b. Furthermore, a 15 sec exposure photograph of high rate gamma events produced by an external
Cs-137 source illuminating the whole THGEM plane is shown in Figure~\ref{cryophotos}c.
 Figure~\ref{LArCCDIntevsEfield} shows the light collection increase with bottom THGEM electric field variation recorded with the CCD camera 
 using 8$\times$8 binning and a 10 sec exposure. 
 The CCD Gaussian mean intensity values reported here are solely for the pixels that contain the alpha source.

 The secondary scintillation light is produced by the passage of electrons through noble gas within a linear 
 electric field~\cite{Monteiro:2007, Monteiro:2008, Phil:2009, Monteiro:2012}
 and as such is expected to increase linearly with the increase in the field up to a point. When a threshold in the electric field is passed the drifting electrons gain enough kinetic energy to ionise the atoms of the medium and subsequently initiate further multiplication known as avalanche, therefore there is an exponential relationship between charge multiplication (and so light) and electric field.
 As shown in Figure~\ref{LArCCDIntevsEfield} the light increases exponentially with electric field indicating that we are within the avalanche region
 however, the gain based on the source could not be determined due to the preamplifier noise.
 A calculated maximum gain estimate of 45 can be made since we know that $\approx$~4000 Am-241 induced electrons reach the extraction region (i.e. after losses due to recombination) 
 and the minimum amount of detectable charge determined from the gas measurements detailed in Section~\ref{gastest} is 175000 electrons.


\begin{table}
\caption[Electric Field Configuration] {Configuration of the electric fields applied in two phase operation.}
\begin{minipage}{\textwidth}
\begin{center}
\vspace{4mm}
{\footnotesize \begin{tabular}{lcccc}
\toprule
 & \minitab[c]{Distance to the \\ stage above (cm)} 
& Potential (kV) & \minitab[c]{Field to the stage \\ above (kV/cm)}\\
\midrule
THGEM$_{2}$ (top electrode) & - & + 2.1 & -  \\
THGEM$_{2}$ (bottom electrode) & 0.1 & + 0.1 & 20  \\
THGEM$_{1}$ (top electrode) & 0.4 & 0 & 2.5 \\
THGEM$_{1}$ (bottom electrode) & 0.1 & -3.8 to -4.1 & 38 to 41  \\
Extraction grid & 1.0 & -10  &
 \minitab[c]{7.08 to 7.44 (gas)\footnote{The field in gas is 1.5 times stronger than in liquid due to the change of the dielectric constant.}\\4.72 to 4.96 (liquid)} \\
Cathode & 20 & -20  & 0.5  \\
\bottomrule
\end{tabular}}
\end{center}
\end{minipage}
\label{LAr_Efields}
\end{table}

\section{Summary \& Conclusions}

The focus of this paper has been the optical imaging of the secondary scintillation light generated in THGEM holes using a CCD 
camera within a two-phase cryogenic LAr TPC. Within this framework we have also developed an original high voltage feedthrough and
we have assessed the use of  web cameras as a monitoring tool within our cryogenic TPC. 

The successful operation of the 
 novel LAr insulated high voltage feedthrough represents a feasible and scalable option for larger detectors.
 The limitations of this design are the specification of the HV coaxial cable and the breakdown voltage of the alumina ceramic 
 (24.5 to 37.0~kV/mm depending on purity and surface finish~\cite{Owate:1992conference, Owate:1992}).
  
 The Microsoft HD-3000 webcam (model no:1456) was found to be the superior option of all the webcams tested in LAr and provided a very useful internal detector
 monitoring tool, allowing close observation of the LAr level during filling. Furthermore, the insight into the internal workings of the detector revealed that the LAr level
 remains constant and steady during recirculation allowing data collection whilst the pump was on.
 
 Characterisation of the Sony ICX285AL CCD chip in a cryogenic environment revealed a lowest operating temperature of $-120$~$\degr$C.
 To overcome this problem a heater was mounted on the back of the chip. There are however, more expensive alternative light sensitive chips that are guaranteed to operate
 to $-200$~$\degr$C such as those manufactured by e2v. In this setup the majority of the electronics such as the digitiser were mounted outside of the detector connected via a
 custom made Kapton cable thus 
 limiting the components required to function in cryogenics to the chip. 
 The VUV secondary scintillation light produced in the THGEM holes was converted to visible with a TPB coated perspex disk placed above the THGEMs 
 allowing the use of economical conventional lenses.
 
 The THGEMs and CCD camera performance was evaluated in argon gas ambient temperature. 
 The highest THGEM gain reached was approximately 1000 
 and for such high gain individual alpha tracks were identifiable with a 1~msec exposure. When 5~sec exposures were taken the overall light was enough
 to light up the individual 500~$\mu$m THGEM holes.

 In two phase conditions accurate determination of the THGEM gain was not possible as the charge signal could not be separated from the preamplifier
 noise. However, for 10~sec exposures photographs of the secondary scintillation light produced by the Am-241 source in LAr were successfully captured. The light detected by 
 the CCD was found to have an exponential increase with the THGEM electric field.
 
 Now that we have demonstrated proof
 of concept, the next stage will be to investigate the capabilities of more light sensitive and ultra fast camera systems that would ultimately allow the
 time resolution of tracks. CCD chips are limited by a readout time of a few msecs however, some state of the art CMOS chip based cameras
 can record 10~$\mu$s exposures with 2~$\mu$s dead time between frames. 
 It is likely that a custom system would need to be developed that allows for the electronics to
 be separated from the chip whilst still maintaining the high readout rate.


\acknowledgments

The Authors are grateful for the expertise and dedicated contributions of the Mechanical Workshop of the Physics
Department, University of Liverpool. We also thank John Bland from the HEP IT group for his support.
We acknowledge the support of the University of Liverpool and STFC.



\end{document}